# Pushing the limits of magnetic anisotropy in the Sm–Co system

S.-Y. Jekal,[1,2,*] M. Charilaou,[1,†] and J. F. Löffler[1]

[1]*Laboratory of Metal Physics and Technology, Department of Materials, ETH Zurich, 8093 Zurich, Switzerland*
[2]*Condensed Matter Theory Group, Paul Scherrer Institute, CH-5232 Villigen PSI, Switzerland*
(Dated: July 24, 2018)

Materials based on the Sm–Co system exhibit remarkable magnetic performance due to their high Curie temperature, large saturation magnetization, and strong magnetic anisotropy, which are the result of the electronic structure in Co and Sm and their arrangement in the hexagonal lattice. In this paper we show, using first-principles calculations, mean-field theory, and atomistic Monte Carlo simulations, that slight modifications of the $SmCo_5$ crystal structure, induced by strain or partial substitution of Sm by Ce, change the exchange interaction and increase the magnetocrystalline anisotropy energy drastically. This finding shows how small changes in local-structure environments can generate substantial changes in macroscopic properties and thus enable the optimization of high-performance materials via tailoring at the atomic level.

## I. INTRODUCTION

The magnetism in intermetallic phases of rare-earth (RE) metals and transition metals (TM), such as the high-performance magnet $SmCo_5$, is a result of the synergy between the $4f$ RE electrons, which provide a high anisotropy due to spin-orbit coupling, and the $3d$ TM electrons, which have large magnetic moments and provide strong ferromagnetic exchange interactions, thus enabling long-range order[1,2]. The synergy between $3d$ and $4f$ electrons depends crucially on the local atomic environments, and thus the net performance of a magnet is extremely sensitive to changes of the lattice.

In this paper we show that slight modifications of the local atomic arrangements in $SmCo_5$ generate drastic changes of the magnetic performance, particularly of the magnetocrystalline anisotropy (MCA). Specifically, we will discuss, based on first-principle electronic-structure calculations and atomistic Monte Carlo simulations, that tensile strain in $SmCo_5$ leads to enhanced magnetic anisotropy. Hence, we propose a special case where partial substitution of Sm with Ce generates a material with magnetic performance which is superior to that of the well-known $SmCo_5$.

The permanent magnet $SmCo_5$ has been intensively studied over the past few decades because it exhibits strong uniaxial MCA, large saturation magnetization ($M_s$), and high Curie temperature ($T_C$) of about 1000 K[2–10]. The high magnetic performance of this system, particularly its spectacularly strong MCA, makes it a key component in various precision-sensitive applications. In fact, out of all Co- and Fe-based RE-containing compounds $SmCo_5$ has the largest magnetocrystalline-anisotropy energy $E_{MCA}$[2,11]. It is thus of fundamental interest to investigate the limits of magnetism in this high-performance system and to find ways of how to enhance the magnetic anisotropies beyond the limit of this well-known compound.

Experimental and theoretical studies have reported possibilities of improving the magnetic performance of $SmCo_5$ both in bulk materials[12,13] and in thin-film structures[14–17]. Promising approaches to enhance the magnetic performance are e.g. the partial substitution of Co with other transition metals (TM), such as Cu, Ni, Fe, or Zr in $Sm(Co, TM)_5$[17–19], but in most cases the $T_C$, the saturation magnetization ($M_s$), and the $E_{MCA}$ decrease with increasing TM content due to the reduced exchange coupling parameters when Co is substituted by other elements[17,20].

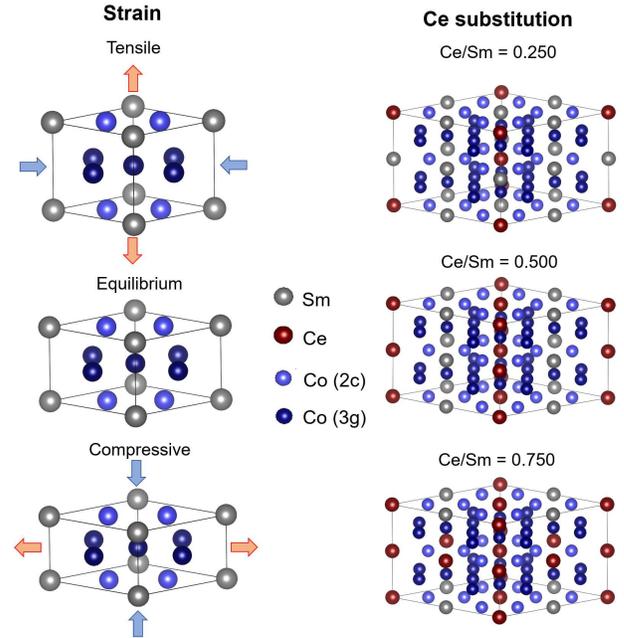

FIG. 1. Illustration of the crystal structure of $SmCo_5$ and $(Sm,Ce)Co_5$. In the left panel, the ideal hexagonal $CaCu_5$ prototype structure with and without strain is shown, and tensile and compressive strains are indicated by red and blue arrows, respectively. The right panel shows the crystal structure of $(Sm_{1-x}Ce_x)Co_5$ for $x = 0.25$, $0.50$ and $0.75$. Gray, red, light blue and dark blue spheres correspond to Sm, Ce, Co(2c) and Co(3g) atoms, respectively. Note that these are the most stable structures obtained by total-energy minimization in the DFT calculations.

Considering the above, the question remains of whether or not it is possible to increase the $E_{MCA}$, and thus enhance the performance of Sm–Co without substituting Co, but by carefully modifying the local atomic arrangements, i.e., by introducing strain. Strain is a particularly important issue in permanent magnets, because their synthesis typically involves sintering, and thermal processes inevitably introduce strain on fine-grained materials[21,22]. Strain can be applied externally by mechanical means, but it can also be introduced to the system by partial element substitution, such as e.g. of Sm by another RE metal with a different atomic radius.

Ce is a particularly promising candidate for substituting Sm because it is the most abundant and second-lightest among all RE metals, and CeCo$_5$ is isomorhpous to SmCo$_5$[23,24], suggesting that it is possible to substitute Sm with Ce through the entire composition, i.e., (Sm$_{1-x}$Ce$_x$)Co$_5$ for $0 \leq x \leq 1$. SmCo$_5$ and CeCo$_5$ crystallize in the hexagonal CaCu$_5$ structure, where each Sm (Ce) occupies a 1a site and Co occupies 2c and 3g sites in the lattice[25] (see Fig. 1); the unit cell volume decreases from 84.71 Å$^3$ in SmCo$_5$ to 83.09 Å$^3$ in CeCo$_5$.

It is well-known that the intrinsic magnetic material parameters of CeCo$_5$ lead to a magnetic performance that is substantially weaker than that of SmCo$_5$[11], i.e., reduced $T_C$ of 660 K instead of 1000 K, reduced $E_{MCA}$ of 5.3 MJ/m$^3$ instead of 17.2 MJ/m$^3$, and reduced $M_s$ of 0.95 T instead of 1.15 T. Nevertheless, a *partial* substitution of Sm by Ce can have drastic effects on the magnetic performance, as we discuss below, because it introduces strain in the structure and breaks the lattice symmetry in a way that enhances the contribution of the Co atoms to the MCA.

In the following we briefly discuss the theory and the computational methodology and then analyze our results, which reveal a surprisingly enhanced magnetic performance of SmCo$_5$ with strain and with partial Sm substitution.

## II. COMPUTATIONAL METHOD

Density Functional Theory (DFT) calculations were mainly performed using plane wave basis sets and pseudopotentials, implemented in Quantum Espresso (QE)[26]. This approximation enables time-efficient calculations, but its precision in certain cases might be limited by the use of pseudopotentials. Hence, results from QE were complemented by more computationally demanding all-electron full-potential augmented plane-wave and local-orbitals basis sets, implemented in Elk[27] and WIEN2k[28], respectively, which enable more precise calculations. This is particularly important for the precise estimation of the MCA energy, which is associated with variations on the order of meV.

All DFT calculations were performed on a $2 \times 2 \times 2$ supercell of the SmCo$_5$ (CaCu$_5$) structure. For the exchange-correlation potential we adapted the local spin-density approximation plus Hubbard $U$ (LSDA+$U$), which can adequately describe the strongly correlated electronic states of $4f$ electrons[9,29,30]. The LSDA+$U$ method requires the Coulomb energy ($U$) and the exchange energy ($\mathcal{J}$) as input parameters, and we defined $U$ and $\mathcal{J}$ through the derivatives of the energy levels $\epsilon_f$ of the $f$-orbital with respect to their occupancies $n_f$, described as $U = \partial \epsilon_{f\uparrow}/\partial n_{f\downarrow}$ and $\mathcal{J} = \partial(\epsilon_{f\uparrow} - \epsilon_{f\downarrow})/\partial(n_{f\uparrow} - n_{f\downarrow})$ for the majority (minority) spin $\uparrow$ ($\downarrow$), respectively. From these expressions we obtain $U = 6.0$ eV and $\mathcal{J} = 1.0$ eV for Sm in our DFT calculations.

Wave functions were expanded by a plane-wave basis set with an optimized cutoff energy of 340 Ry, and the Brillouin zone was sampled via a $12 \times 12 \times 15$ k-point mesh. Different mesh values were tested to ensure the numerical stability of our calculations, with the convergence criterion being 0.1 µeV. All structures were fully relaxed prior to calculating the magnetic parameters, i.e., the atomic arrangements in our calculations correspond to the most stable structures, found by minimizing the total energy.

The magnetocrystalline anisotropy energy $E_{MCA}$ was calculated using the force theorem and is defined as the total energy difference between the magnetization perpendicular to the c-plane and in the c-plane of the SmCo$_5$ structure, i.e., $K = E_{[100]} - E_{[001]}$, where $E_{[100]}$ and $E_{[001]}$ are the total energies with the magnetization aligned along the hard- ([100]) and easy-axis ([001]) of the magnetic anisotropy, respectively, with (001) corresponding to the c plane. Specifically, $E_{MCA}$ is calculated in three steps: (i) collinear self-consistent calculation without spin-orbit coupling (SOC); (ii) global rotation of the density matrix to consider the magnetization along [100] and [001] to calculate $E_{[100]}$ and $E_{[001]}$, respectively; and (iii) non-collinear and non-self-consistent calculation with SOC.

For the calculation of the energy-resolved distribution of the orbitals within second-order perturbation theory[31], $E_{MCA}$ can be expressed as

$$E_{MCA}^{\uparrow\downarrow} \propto \xi^2 \times$$
$$\sum_{o^{\uparrow(\downarrow)}, u^{\downarrow(\uparrow)}} \frac{|<o^{\uparrow(\downarrow)}|L_x|u^{\downarrow(\uparrow)}>|^2 - |<o^{\uparrow(\downarrow)}|L_z|u^{\downarrow(\uparrow)}>|^2}{\epsilon_{o^{\uparrow(\downarrow)}} - \epsilon_{u^{\downarrow(\uparrow)}}}, \quad (1)$$

where $\xi$ is the spin-orbit coupling constant, $\uparrow(\downarrow)$ represent majority (minority) spins, and $o(u)$ and $\epsilon_{o(u)}$ represent the eigenstate and eigenvalue of occupied (unoccupied) states, respectively. For contributions from the $d$-orbitals, the matrix elements of the $L_x$ and $L_z$ operators can be expressed via the magnetic quantum number $m$. When the occupied and unoccupied states have opposite spin directions, $m_o - m_u = \pm 1$ and $m_o - m_u = 0$ contribute to out-of-plane and in-plane MCA, respectively, while $m_o - m_u = \pm 2$ is treated as zero matrix element.

From the DFT calculations we obtain the magnetocrystalline anisotropy energy, as described above, and from the total energy minimization we obtain the atomic magnetic moments and the inter-atomic ferromagnetic exchange interactions as the difference between ferromagnetic and antiferromagnetic spin configuration, i.e., $J = (E_{\uparrow\uparrow} - E_{\uparrow\downarrow})/2$. We then use these intrinsic material parameters as inputs for atomistic Monte Carlo simulations to calculate finite-temperature material properties by mapping the material parameters onto a Heisenberg-type model. Here, the Hamiltonian has the form

$$\mathcal{H} = -\frac{1}{2} \sum_{i \neq j} J_{ij} \vec{S}_i \cdot \vec{S}_j - \sum_j K_i (S_z)^2 , \qquad (2)$$

where the spin $\vec{S}_i$ is the 3-dimensional vector at each lattice site, $J_{ij}$ is the Heisenberg exchange interaction energy between neighboring sites, and $K_i$ is the site-specific uniaxial anisotropy energy constant. The numerical values of the interaction and anisotropy energies were taken directly from the DFT results in absolute values.

The simulations were performed by means of the Metropolis algorithm, i.e., by single-site updates. For the thermalization the algorithm was run 5000 MCS (Monte Carlo steps per site) and another 5000 MCS were run to sample the thermal-average magnetic moment $\vec{M}$ at the 1a, 2c, and 3g lattice sites at each temperature.

We considered systems with 120000 and 276000 spins, corresponding to real dimensions of 5 nm × 8.5 nm × 39 nm and 10 nm × 10 nm × 40 nm, respectively, to verify the numerical stability of our results. Additionally, we used periodic boundary conditions to eliminate surface effects.

For the calculation of the Curie temperature we started the simulations with all spins parallel and pointing along [001] at $T = 0$ and gradually increased the temperature to 1200 K, whereas for the calculation of the finite-temperature domain-wall thickness, we started the simulations with half the spins pointing along [001] and the other half pointing along [00$\bar{1}$] and thermalized the system at each given temperature, in a similar way as described in Ref.[32].

## III. RESULTS AND DISCUSSION

### A. Strain effects in SmCo$_5$

We calculated the total energy of the SmCo$_5$ structure to optimize the lattice and determined the in-plane lattice constant to be $a = 4.980$ Å and the axis ratio to be $c/a = 0.792$. These values are in excellent agreement with previous first-principles calculations[9,29,33,34] and experiments[35], thus confirming our computation of the equilibrium structure.

Further, for SmCo$_5$ in equilibrium we obtain an $E_{\text{MCA}}$ of 9.09 meV/f.u. (f.u. = formula unit), which is again in excellent agreement with experimental data[11,36] (corresponding to 17.18 MJ/m$^3$) and with other DFT calculations[19,33]. Specifically, we find that in the SmCo$_5$ structure the Co sublattice contributes 2.20 meV and the Sm sublattice contributes 6.88 meV to the total anisotropy energy per unit cell (corresponding to 4.16 MJ/m$^3$ and 13.02 MJ/m$^3$, respectively, in good agreement with experimental values for the individual contributions[36]).

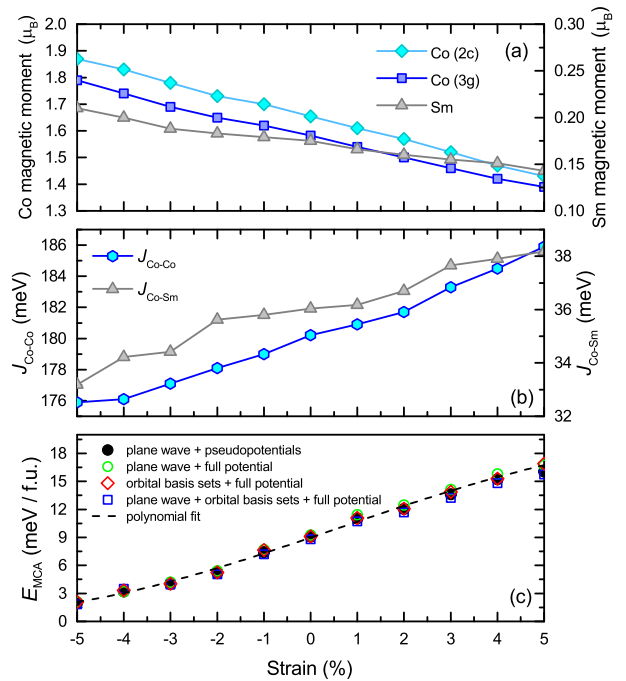

FIG. 2. Intrinsic magnetic properties as a function of strain in SmCo$_5$: (a) atomic magnetic moments of Co atoms at sites 2c (diamonds) and 3g (squares), and Sm atoms (triangles); (b) effective total ferromagnetic exchange energy $J_{\text{Co}-\text{Co}}$ between Co atoms (hexagons) and $J_{\text{Co}-\text{Sm}}$ between Co and Sm atoms (triangles); and (c) magnetocrystalline anisotropy energy ($E_{\text{MCA}}$), calculated by several approximations.

In order to probe SmCo$_5$ away from equilibrium we applied both tensile (+) and compressive (−) strain along the [001] axis (see Fig. 1), while keeping the unit cell volume constant, to investigate strain-dependent electronic structure and magnetic properties.

Strain effects on the magnetism, i.e., individual atomic magnetic moments in SmCo$_5$, effective ferromagnetic exchange energy, and MCA energy are presented in Fig. 2. We find that the magnetic moments at each crystallographic site decrease nearly linearly with increasing strain (see Fig. 2a). The reason for the decrease in the magnetic moments is the decreasing overlap of the d-orbitals as the c axis grows with strain.

Further, we calculated the overall effective exchange coupling parameters in the framework of the multi-sublattice mean-field approximation, and we obtained

$J_{\text{Co-Co}}$ and $J_{\text{Co-Sm}}$, which are the sum of the exchange coupling constants within a sphere of radius $R = 5a$. From these calculations we observe that the total ferromagnetic exchange interaction increases monotonically with increasing strain (see Fig. 2b), in contrast to the magnetic moments. Note that the Curie temperature does not change significantly because the increase in the effective ferromagnetic exchange is offset by the decrease in the magnetic moments.

While the overall exchange interactions increase with increasing strain, the intra-plane interaction strengths $J_{\text{Co(2c)-Co(2c)}}$ and $J_{\text{Co(3g)-Co(3g)}}$ increase while the inter-plane interaction strength $J_{\text{Co(2c)-Co(3g)}}$ decreases with increasing tensile strain (data not shown). This illustrates the dependence of the ferromagnetic exchange on the inter-atomic distance: as the c–c and g–g distances decrease the corresponding exchange interaction increases, whereas as the c–g distance increases with increasing tensile strain the exchange interaction decreases.

Similar strain effects were calculated for YCo$_5$ in Ref.[29,33], but with significantly weaker impact. It was, however, pointed out that the inter- vs intra-plane hybridization plays a key role in the development of the ferromagnetic exchange, the magnetic moments, and $E_{\text{MCA}}$.

Considering the effects of strain on $E_{\text{MCA}}$ in SmCo$_5$, we find that the magnetocrystalline anisotropy energy changes monotonically with strain. While compressive strain decreases $E_{\text{MCA}}$, increasing tensile strain has the opposite effect and $E_{\text{MCA}}$ along the c-axis exhibits a remarkable strengthening (see Fig. 2c). In fact, even with only +1% strain $E_{\text{MCA}}$ increases by $\sim 30\%$, whereas with +5% strain $E_{\text{MCA}}$ increases, strikingly, by 80% compared to that in equilibrium at zero strain. Importantly, we find that the results based on the use of pseudopotentials are in excellent agreement with full-electron calculations, and they only start to deviate by 3–5% with increasing strain, as the values of $E_{\text{MCA}}$ become larger. Specifically, calculations based on orbital basis-sets and plane-waves consistently tend to slightly underestimate $E_{\text{MCA}}$.

We examined the cause of this remarkable enhancement of $E_{\text{MCA}}$ and found that the major contribution comes from the Co(2c) sites. Note that the 2c sites also play a vital role in the MCA of the SmCo$_5$ in equilibrium, as indicated by nuclear magnetic resonance studies[37], and that calculations of strained YCo$_5$ also indicated the importance of 2c contributions to the anisotropy[33].

To obtain a deeper understanding of why the lattice distortion affects the MCA, we plot tensile and compressive strain-induced changes in the electronic density of states (DOS), i.e., $\Delta \text{DOS} = \text{DOS}(+5\%) – \text{DOS}(0\%)$ (Fig. 3a) and $\Delta \text{DOS} = \text{DOS}(-5\%) – \text{DOS}(0\%)$ (Fig. 3b). We observe that, especially in the range near the Fermi level, the changes in the DOS with $m$ (with up and down spins) are mirrored depending on strain directions, i.e., the $m = 0, \downarrow$ and $m = \pm 2, \downarrow$ states lose electrons while the $m = \pm 1, \uparrow$ state gains electrons with tensile strain, and vice versa with compressive strain.

The concurrence of the increased occupied $m = \pm 1, \uparrow$

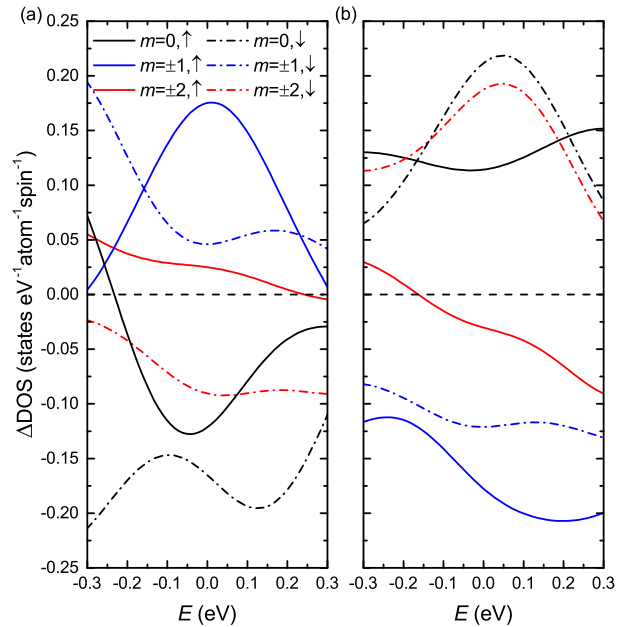

FIG. 3. Changes in DOS of the Co(2c) atom induced by (a) +5% (tensile) and (b) −5% (compressive) strain. Positive change indicates gain of electrons, whereas negative change indicates loss of electrons for each state. Black, blue, and red lines denote $m = 0$, $m = \pm 1$, and $m = \pm 2$, respectively, while solid and dashed lines represent majority and minority states, respectively. The Fermi level is set to zero.

state and decreased unoccupied $m = 0, \downarrow$ and $m = \pm 2, \downarrow$ states around the Fermi level lead to larger $E_{\text{MCA}}$ because of the stronger matrix elements of $<m = \pm 1, \uparrow | L_x | m = 0, \downarrow>$ and $<m = \pm 1, \uparrow | L_x | m = \pm 2, \downarrow>$. Meanwhile, for the compressive strain shown in Fig. 3(b), this process is reversed and $E_{\text{MCA}}$ decreases.

These results show that tensile stain along the c-direction enhances the effective ferromagnetic exchange energy notably but most importantly it causes a striking increase of the MCA energy. This also suggests that off-equilibrium properties that exceed those of the equilibrium state so drastically may be exploited in applications where materials are mechanically strained, either continuously or in pulses. Importantly these off-equilibrium material parameters provide a deeper understanding of the fundamental mechanisms of magnetic anisotropy and ferromagnetic exchange in RE–TM systems and thus may enable the design of materials with enhanced properties.

Inspired by these findings, we now turn to a case study where we partially substitute Sm by Ce to induce local strain in the cell and to predict changes in magnetic performance.

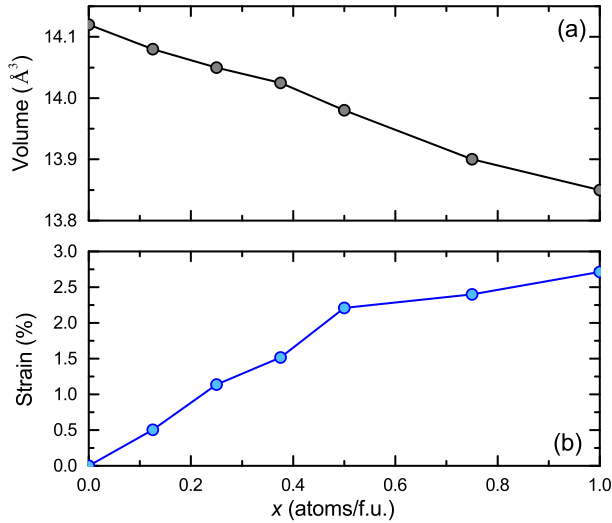

FIG. 4. Change of (a) unit-cell volume and (b) effective strain in $(Sm_{1-x}Ce_x)Co_5$ as a function of Ce concentration $x$.

## B. Partial substitution of Sm by Ce

We performed calculations of the $(Sm_{1-x}Ce_x)Co_5$ ($x$ = Ce/Sm) system, where we considered seven different compositions: $x$ = 0, 0.125, 0.25, 0.375, 0.5, 0.75, and 1. For each $x$ the total energy was minimized to obtain the most stable atomic arrangement among all possible configurations and the corresponding unit-cell volume. Examples of the systems with $x$ = 0.25, 0.5, and 0.75 are shown in Fig. 1, where gray, red, light blue and dark blue spheres correspond to Sm, Ce, Co(2c) and Co(3g) atoms, respectively.

In order to obtain a direct comparison between the effects of Ce-substitution on the magnetic properties with those of strain, we show in Fig. 4(a) and (b) the unit-cell volume and the effective strain, respectively, as function of the Ce concentration $x$. The volume decreases nearly linearly with increasing $x$, which is due to the smaller atomic radius of Ce compared to that of Sm, whereas the strain increases correspondingly monotonically with increasing $x$. This substitution-induced tensile strain can have drastic effects on the magnetic state, which do not necessarily correspond to the electronic properties of Ce that would tend to decrease the anisotropy energy.

The calculated magnetic moments of the Co, Ce, and Sm atoms decrease monotonically with increasing Ce-substitution $x$, which correlates with the decreasing atomic volume, following Vergard's law[38] (see Fig. 5a). Also, the calculated $J_{Co-Co}$ and $J_{Co-RE}$, which is the averaged interaction between Co and Sm or Ce atoms, decrease strongly with increasing $x$, as seen in Fig. 5(b).

For the magnetocrystalline anisotropy, however, we find a surprising non-monotonic behavior. In fact, for $x$ = 0.125 the anisotropy energy increases to 10.49 meV/f.u. (19.8 MJ/m$^3$) and for $x$ = 0.25 it becomes 11.63 meV/f.u. (22.0 MJ/m$^3$), which is 30% larger than that of the SmCo$_5$ system in equilibrium (see Fig. 5c). As $x$ increases further, however, $E_{MCA}$ decreases and for $x$ = 1, i.e., CeCo$_5$, it becomes 2.96 meV/f.u. (corresponding to 5.6 MJ/m$^3$). This $E_{MCA}$ for CeCo$_5$ is also in excellent agreement with experimental data[11].

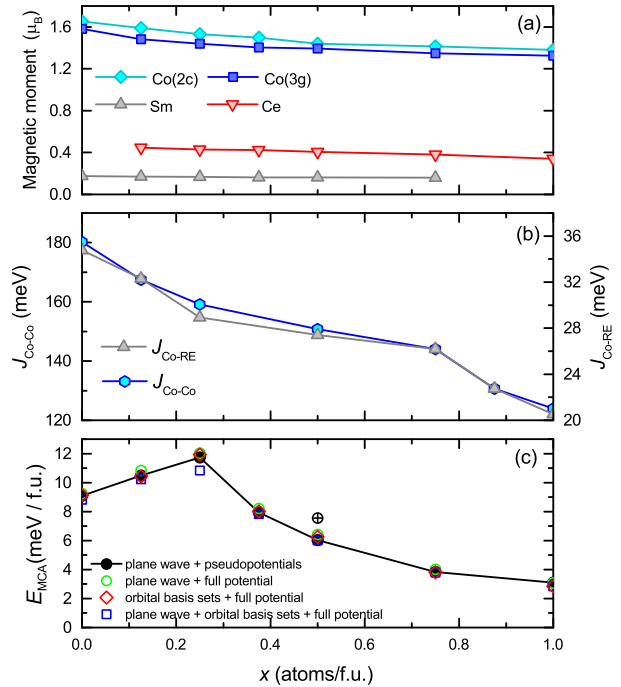

FIG. 5. Magnetic properties as a function of Sm-substitution by Ce: (a) atomic magnetic moments for Co on sites 2c (diamonds) and 3g (squares), Sm (triangles), and Ce (inverse triangles); (b) effective total ferromagnetic exchange interaction energy $J_{Co-Co}$ between Co atoms (hexagons) and $J_{Co-RE}$ between Co and RE atoms (triangles), averaged over Co–Sm and Co–Ce pairs; and (c) magnetocrystalline anisotropy energy ($E_{MCA}$). The size of the symbols is larger than the error bars. In panel (c) the crossed circle corresponds to the $E_{MCA}$ of metastable systems with $x$ = 0.5.

Note that for $x$ = 0.5 the equilibrium structure has $E_{MCA}$ = 6.18 meV/f.u., which corresponds to only 68% of that of the SmCo$_5$ compound, but we found metastable configurations that yield a magnetic anisotropy energy of 7.56 meV/f.u., which is only slightly smaller than that of SmCo$_5$ despite the fact that half of the Sm is substituted by Ce.

The 30% enhancement in $E_{MCA}$ for $x$ = 0.25 is unexpected, given that the electronic structure of Ce and its contribution would tend to substantially lower the anisotropy. We find, however, that this increase correlates with the strain in the unit-cell of SmCo$_5$ induced by the substitution of every 4th Sm atom by a Ce atom, which corresponds to 1.13% strain (see Fig. 4b). For the non-substituted SmCo$_5$ system, 1% of strain corresponds to 22% increase in $E_{MCA}$, as shown in Fig. 2c.

Hence, this means that with 1/4 Ce substitution the effects of the strained Sm–Co structure dominate over the electronic contributions of Ce.

Again we find excellent agreement between the different approximations, with an exception at $x = 0.25$, which exhibits the highest $E_{MCA}$ and is slightly underestimated by the approximation based on orbital basis-sets and plane waves.

The main contribution to the enhancement of $E_{MCA}$ comes from the Co(2c) sites, which is in direct analogy to the effects of strain (see Fig. 3). In order to investigate in detail the origin of the enhanced $E_{MCA}$ of $(Sm_{0.75}Ce_{0.25})Co_5$ and the decreased value in high Ce concentration cases, we present in Fig. 6 the DOS decomposed into $d_{xy,x^2-y^2}$, $d_{yz,xz}$ and $d_{z^2}$ orbitals, which correspond to magnetic quantum numbers of $m = \pm 2$, $m = \pm 1$, and $m = 0$, respectively. To provide a clear explanation, only the DOS of Co(2c) atoms that drive the change of $E_{MCA}$ is presented.

system with $(Sm_{0.75}Ce_{0.25})Co_5$ the unoccupied minority $m = 0$ and $m = \pm 2$ orbitals, indicated by arrows, shift toward the Fermi level. As a result, $E_{MCA}$ is enhanced because the $< m = \pm 1, \uparrow |\ L_x\ |\ m = 0, \downarrow >$ and the $< m = \pm 1, \uparrow |\ L_x\ |\ m = \pm 2, \downarrow >$ coupling become stronger. For the $CeCo_5$ case, however, the peaks of minority $m = 0$ and $m = \pm 2$ orbitals move to the occupied states, and therefore couplings which contribute to the $E_{MCA}$ are strongly weakened.

Given the above, first-principles calculations provide compelling evidence that strain in $SmCo_5$, either in the form of mechanical strain or due to partial Sm-substitution by Ce in $(Sm_{1-x}Ce_x)Co_5$, can push the limits of magnetic anisotropy in this system. The agreement between our results and existing literature for the equilibrium properties of $SmCo_5$, both structural and magnetic, confirms our calculations and makes our predictions for the strained structures very pron...

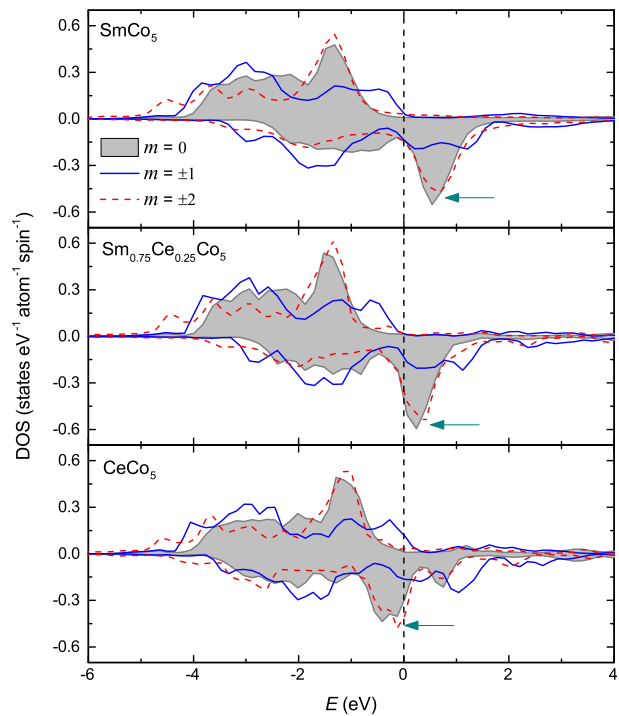

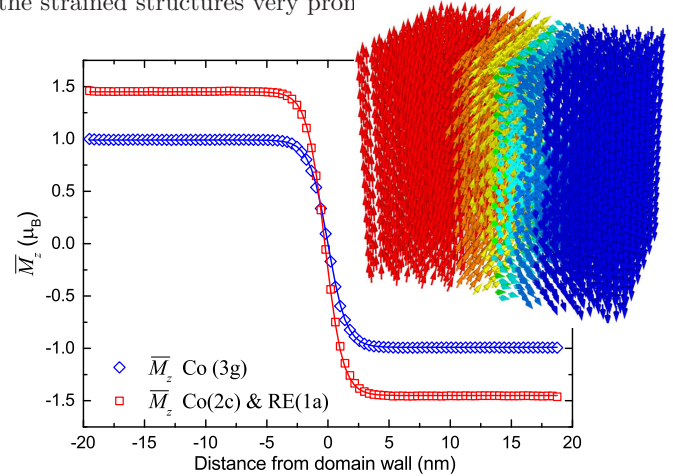

FIG. 6. Comparison of the $d$-projected DOS of Co(2c) atoms in $SmCo_5$, $(Sm_{0.75}Ce_{0.25})Co_5$ and $CeCo_5$. Shaded, solid, and dotted lines denote $m = 0$, $m = \pm 1$, and $m = \pm 2$, respectively. The Fermi level is set to zero. Arrows emphasize orbital shifting.

FIG. 7. Monte Carlo simulation at finite temperature showing the domain-wall profile of the sublattice containing Co(3g) atoms and the sublattice containing RE(1a) and Co(2c) atoms. A snapshot of the system containing a domain wall is presented at the corner.

As shown in Fig. 6(a), in the DOS of $SmCo_5$ we find $< m = \pm 1, \uparrow |\ L_x\ |\ m = 0, \downarrow >$ and $< m = \pm 1, \uparrow |\ L_x\ |\ m = \pm 2, \downarrow >$ couplings, as the peaks of $m = \pm 1$ and $m = 0$ lie directly below and above $E_F$, respectively, and the energy difference between occupied and unoccupied states is reduced (see Eq. 1), leading to the strong MCA along the $c$ axis. In the substituted

All calculated material parameters so far correspond to the intrinsic magnetic properties at zero temperature. Hence, we extend now our computational analysis to finite temperature to predict both the intrinsic material parameters and the resulting relevant length scales, i.e., the domain-wall thickness $\delta_{dw}$, of Ce-substituted Sm–Co materials. The computation of finite-temperature length scales takes into account the effects of the microstructure and is thus particularly relevant to high-temperature applications.

Taking the DFT results shown in Fig. 5 as inputs, we performed Monte Carlo simulations and computed the length scale of a domain wall by simulating domain-wall behavior in the Sm–Ce–Co system. An example is illustrated in Fig. 7, which shows the spatially-resolved sublattice magnetization $\overline{M}_z$ as a function of the distance from the domain-wall center. The spins rotate around the center of the domain wall and the profile is proportional to $\tanh(r/\lambda)$, with $r$ the distance from the domain-wall center and $\lambda = \delta_{\rm dw}/2$. Note that we distinguish between the magnetization of the 3g sites and the 1a and 2c sites.

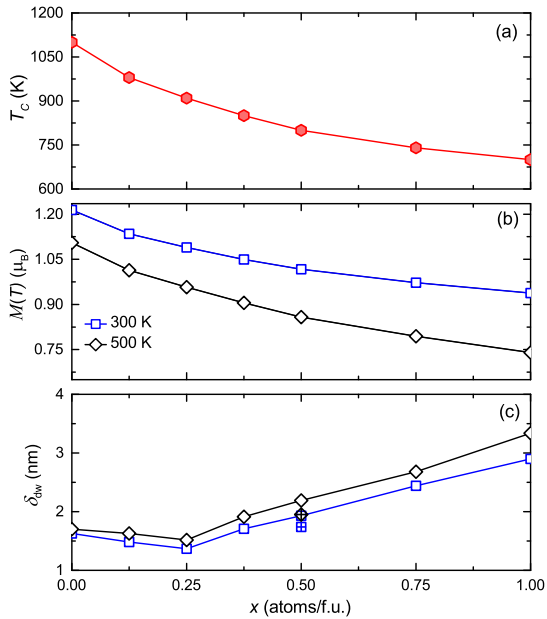

FIG. 8. Effect of the Ce-substitution on the finite-temperature magnetic properties: (a) Curie temperature; (b) average magnetic moments at $T = 300$ K and 500 K; and (c) domain-wall thickness at $T = 300$ K and 500 K, as a function of the Ce concentration $x$. The size of the symbols is larger than the error bar.

In order to obtain a comparison between all the Ce-substituted systems, we computed the Curie Temperature $T_{\rm C}$, the average atomic magnetic moment $M$, and $\delta_{\rm dw}$ at $T = 300$ K and 500 K ($= 0.5 T_{\rm C}^{\rm SmCo_5}$). Figure 8 shows these results as a function of the Ce concentration.

Both $T_{\rm C}$ and $M$ gradually decrease with increasing Ce concentration. For SmCo$_5$ ($x = 0$) we obtain $T_{\rm C} \approx 1100$ K, whereas for CeCo$_5$ ($x = 1$) we obtain $T_{\rm C} \approx 700$ K, both of which are in good agreement with experimental data ($\sim 1050$ K for SmCo$_5$ and $\sim 660$ K for CeCo$_5$).

For the domain-wall thicknesses, we find that for SmCo$_5$ at room temperature $\delta_{\rm dw} = 1.61(1)$ nm, which is in very good agreement with previous theoretical calculations[39] ($\delta_{\rm dw} = 1.6$ nm) and experiments[40] ($\delta_{\rm dw} = 1.3 \pm 0.3$ nm), supporting the validity of our first-principles calculations and MC simulations, considering that SmCo$_5$ without Ce is our reference system.

The $\delta_{\rm dw}$ reflects the tendency of $E_{\rm MCA}$ shown in Fig. 5, with the smallest domain wall being at $x = 0.25$ and the largest at $x = 1$. Note that the decreasing domain-wall thickness in the Ce-substituted system suggests that a domain wall could fit in smaller structures and be more easily trapped, thus potentially increasing the coercivity of a nanostructured magnet, and hence ensuring that the same microstructure that makes SmCo$_5$ a high-performance magnet would also make (Sm$_{0.75}$Ce$_{0.25}$)Co$_5$ a permanent magnet with even higher performance, especially with regards to its MCA, which is the crucial component for precision applications.

## IV. CONCLUSIONS

We found, based on electronic structure calculations, that changes in the lattice spacing of SmCo$_5$ have a strong effect on the density of states close to the Fermi level, causing dramatic changes in the magnetic properties. Specifically, we observed that tensile strain in SmCo$_5$ monotonically enhances the magnetocrystalline anisotropy energy ($E_{\rm MCA}$) and the effective ferromagnetic exchange interaction ($J_{\rm eff}$), while reducing the atomic magnetic moments of Sm and Co. In contrast, compressive strain reduces the $E_{\rm MCA}$ and the $J_{\rm eff}$, and increases the atomic magnetic moments. Further, we investigated the effects of strain induced by partial substitution of Sm by Ce and discovered that it has also drastic effects on the $E_{\rm MCA}$. In fact, when Ce substitutes every 4th Sm atom in the lattice the $E_{\rm MCA}$ increases by more than 30%, compared to that of SmCo$_5$, because the density of states increases close to the Fermi level. In order to predict the potential of this system for high-performance applications, we investigated the finite-temperature properties of Ce-substituted SmCo$_5$ and found that the magnetic length scales, the atomic magnetic moments, and the Curie temperature are comparable to those of the parent compound SmCo$_5$, suggesting that the Ce-substituted material is a very promising candidate for high-performance high-temperature applications with enhanced magnetocrystalline anisotropy and with 25% reduced heavy-rare earth content. Further, we have shown that multi-scale modeling is a powerful strategy for the detailed analysis and prediction of functional materials properties.

## V. ACKNOWLEDGMENTS

The authors acknowledge funding from the Swiss National Science Foundation via Grant No. 200021–165527 and from the ETH Grant ETH-47 17-1, and thank P. M. Derlet for fruitful discussions.